\documentclass[aps,twocolumn,prb,tightenlines,floatfix,showpacs]{revtex4}
\usepackage{graphicx}
\usepackage[english]{babel}
\usepackage{amsmath}
\usepackage{amssymb}

\newcommand{\fk}{f_{\mathbf{k}}}

\newcommand{\vk}{\mathbf{k}}

\newcommand{\be}{\begin{eqnarray}}
\newcommand{\ee}{\end{eqnarray}}

\newcommand{\ad}{a^{\dagger}}
\newcommand{\bd}{b^{\dagger}}
\newcommand{\cd}{c^{\dagger}}
\newcommand{\dd}{d^{\dagger}}

\newcommand{\vq}{\mathbf{q}}

\newcommand{\vR}{\mathbf{R}}

\newcommand{\tsigma}{\tilde{\sigma}}

\def\ket#1{|#1\rangle}
\def\bra#1{\langle #1 |}

\begin{document}

\title{Collective Modes in the Loop Current Ordered Phase of Cuprates}

\author{Yan He and C.M. Varma}

\affiliation{Department of Physics, University of California, Riverside, CA}
\date{\today}
\begin{abstract}

Recently two branches of weakly dispersive collective modes have been discovered in under-doped cuprates by inelastic neutron scattering. Polarization analysis reveals that the modes are magnetic excitations. They are only visible for temperatures below
the transition temperature to a broken symmetry phase which was discovered earlier and their intensity increases as temperature is further decreased. The broken symmetry phase itself has symmetries consistent with ordering of orbital current loops within a unit-cell without breaking translational symmetry. In order to calculate the collective modes of such a state we add quantum terms to the Ashkin-Teller (AT) model with which the classical loop current order has been described. We derive that the mean field ground state of the quantum model is a product over all unit-cells of linear combination of the four possible classical configurations of the loop current order in each unit-cell. The collective modes  are calculated by using a generalized Holstein-Primakoff boson representation of orbital moment operators and lead to three branches of gapped weakly dispersive collective modes. The experimental results are consistent with the two lower energy branches; the third mode is at  a higher energy than looked for by present neutron scattering experiments and might also be over-damped. Implications of the discovery of the collective modes are discussed.

\end{abstract}

\maketitle

\section{introduction}
\begin{figure}
\centerline{\includegraphics[width=0.5\textwidth]{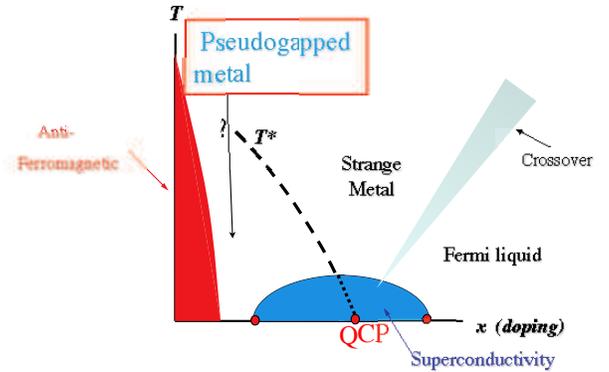}}
\caption{The universal phase diagram of hole-doped cuprates based on properties which show characteristic changes across the boundaries shown in all the cuprates. The boundary of the pseudogap phase has not yet been determined in experiments, i.e how $T^*(x)$ line continues for lower $x$ has not been determined.}
\label{phase-diagram}
\end{figure}

Thermodynamic \cite{pseudogap-susc, loram-spht.} as well as other properties \cite{norman-pines-kallin} in all  Cuprate families with high superconducting transition temperatures change their temperature dependence below about a characteristic temperature $T^*(x)$, which depends on doping $x$, from those at higher temperatures.  The region below $T^*(x)$, see Fig. (\ref{phase-diagram})  is said to mark the pseudogap region.  Photoemission \cite{fisher-stm} and Angle resolved photoemission experiments reveal \cite{pseudogap-arpes}  an anisotropic decrease in single particle spectral weight near the chemical potential below $T^*(x)$. A popular belief \cite{norman-pines-kallin} has been that $T^*(x)$ marks a crossover to a region of reduced low energy one-particle spectral weight, as well as in multiple particle-hole spectral weights, due to one or other reason, preformed pairs \cite{pseudogap-prepair}, resonating valence bonds \cite{pseudogap-rvb}, stripe formation \cite{pseudogap-stripes} or other states of charge modulation, proximity to the Mott-AFM-insulating state \cite{pseudogap-afm}, etc.. Guided by the fact that a ``strange -metal " region whose properties can be explained by quantum-critical fluctuations \cite{mfl} abuts the pseudo-gap region in the phase diagram, a broken symmetry state was sought for the region below $T^*(x)$ with  $T^*(x) \to 0$ at $x \to x_c$, the quantum-critical point. An unusual class of phases \cite{loop-order, cmv-2006} was found to be stable in mean-field calculations of the three-band model for cuprates  \cite{vsa, emery}. In such phases time-reversal symmetry is broken via spontaneous generation of orbital currents in each cu-o unit-cell without altering the translational symmetry. The phase transition belongs to the class with an order parameter singularity but no divergence in the specific heat \cite{cmv-2006, sudbo}.

\begin{figure*}[t]
\centerline{\includegraphics[width=0.9\textwidth]{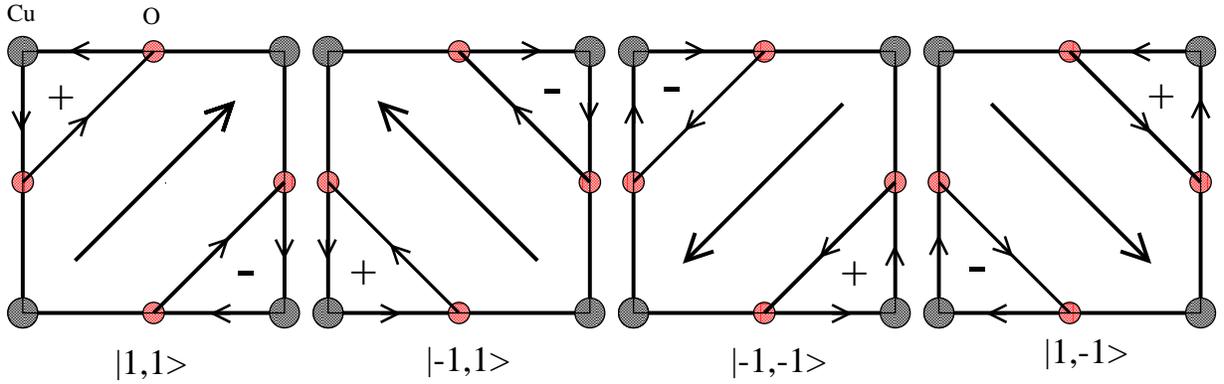}}
\caption{The four Possible ``classical" domains of the loop ordered state are shown. In the classical ordered phase, one of these configurations is found in every unit-cell.}
\label{configs}
\end{figure*}

An order parameter consistent with the symmetry predicted has been discovered in four different families of cuprates through polarized neutron scattering \cite{loop-order-expt} or dichroic ARPES \cite{kaminski} and may be regarded as a universal property of the cuprates.  The magnitude of the order parameter is large, estimated in $Hg{1201}$, with $T_c \approx 61 K$ and $T^* \approx 350 K$, to be about 0.4$\mu_B$ per unit-cell.

More recently inelastic neutron scattering  has discovered \cite{coll-modes} {\it two} branches of weakly dispersive collective modes for three different $x$ in the family ($Hg1201$) for $T$ below $T^*(x)$. (Similar modes have been also found \cite{coll-modes-mook} in $YBa_2Cu_3O_{6.6}$.) Their onset temperature and increase of spectral weight as a function of temperature  follows the temperature of the intensity of the elastic order.  Such  modes were also anticipated  \cite{cmv-2006} and further substantiate the broken symmetry. In this paper, we describe details of the calculation of such modes as well as of another higher energy branch of excitations which has not yet been discovered. A brief report of this work has already been published \cite{he-cmv-prl}.

The observed broken symmetry is consistent with spontaneous moments due to a pair of orbital current loops within each  unit-cell preserving overall translational symmetry. It breaks both time-reversal and inversion symmetry, preserving their product. The ``classical" order parameter \cite{shekhter} may be characterized by the \textit{anapole} vector \cite{zeldovich} ${\bf L}$
\be
\label{def L}
{\bf L} = \int_{cell} d^2r  ({\bf M}({\bf r})\times {\bf \hat{r}}) \approx  \sum_{\mu}{\bf M_{\mu}} \times {{\bf r}}_{\mu}
\ee
where the moment distribution ${\bf M}({\bf r})$ is formed due to the currents on the four O-Cu-O triangles per unit-cell as shown in Fig. (\ref{configs}). This figure also shows the four possible ``classical" domains of the loop current ordered state. In the classical ground state, ordering occurs in one of the domains shown.

\begin{figure}
\centerline{\includegraphics[width=0.5\textwidth]{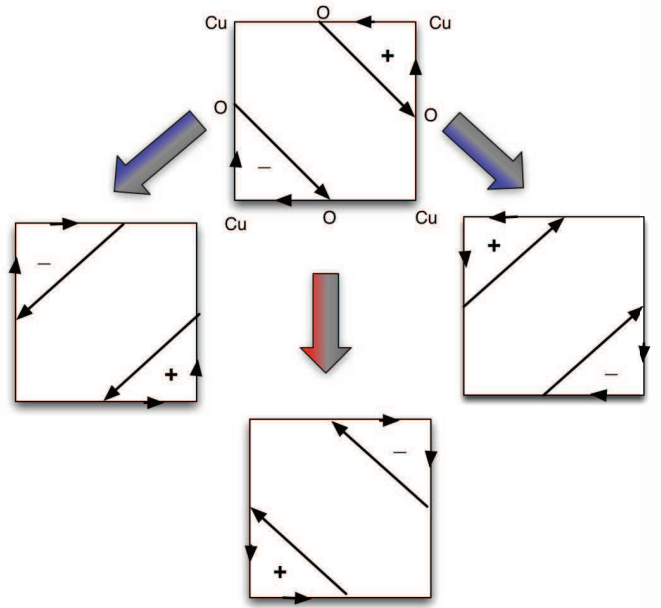}}
\caption{The schematic figure shows that there are only 3 collective modes.}
\label{picture-3 branches}
\end{figure}

Quantum-mechanics allows local fluctuations among the four configurations in Fig (\ref{configs}). This leads, as shown in this paper to a ground state in which each unit-cell has a finite admixture of all the four configurations. It also leads to three branches of  collective modes of the order parameter at finite energies at all momenta ${\bf q}$ for $ T < T^*$. The finite energy follows from the fact that the ground state has symmetry consistent with that of a generalized (transverse-field) Ising model. In this paper these modes will be derived. One can argue that there should be three because each of the four configurations can make transitions to the other three as pictorially shown in Fig. (\ref{picture-3 branches}).

This paper is organized as follows: In the next section, we introduce the classical AT model for the loop current order and generalize it to the quantum model in the $SU(4)$ representation rather than the $SU(2) \times SU(2)$ of the classical AT model. The quantum terms are chosen from considerations of the internal and lattice symmetries of the classical model. In the following section, the ground state of the quantum model is evaluated in mean-field and the dispersion is calculated using the generalization of the Holstein-Primakoff transformation. We compare with the results from experiments. We conclude by discussing the significance of the experimental discovery of the collective modes and the further possible effects which arise from the calculations here. In four Appendices, we discuss the necessity for casting  the problem in the $SU(4)$ representation, some technical details, and the theory for inelastic neutron scattering from the collective modes.

\section{Model for Quantum-Statistical Mechanics of Loop-currents}
\label{sym}

The order parameter ${\bf L}$ and an effective Hamiltonian for this collective variable has been derived \cite{loop-order, cmv-2006, Aji} starting from a model of interacting fermions \cite{vsa}. The full Hamiltonian may be written as a sum of three parts:
\be
H = H_{coll} + H_{F} + H_{F-coll},
\ee
where $H_{coll}$ is the Hamiltonian for the collective co-ordinates ${\bf L}_i$, which order at $T^*$ to give the long range order ${\bf L}$,  $ H_{F}$ is the remnant fermion Hamiltonian and $H_{F-coll}$ is the residual interaction between the fermions and the collective coordinates. In this paper we will be concerned almost exclusively with $H_{coll}$, although in a brief discussion of damping near the end, $H_{F-coll}$ is implicated.

\begin{figure}[t]
\centerline{\includegraphics[width=0.25\textwidth]{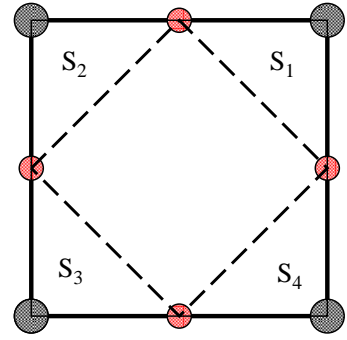}}
\caption{We represent here the notation introduced in the text for the location of the moments within a unit-cell.}
\label{notation}
\end{figure}

An approximate representation of ${\bf L}$ of Eq. (\ref{def L}) is given by ${\bf M}({\bf r})\approx {\bf M_{\mu}} \delta^2({\bf r} -{\bf r}_{\mu})$, where ${\bf r}_{\mu}$, ${\mu} = 1,..,4$ are the location of the four "sites" in any cell at the centroid of the moment distribution. These sites are labeled $S_1,..,S_4$ in Fig.(\ref{notation}). The orbital moments ${\bf M_{\mu}}$ are either up or down or zero. The four {\it classical} domains of Fig.(\ref{configs}) may be represented by the four values of the angle $\theta = \pi/4, 3\pi/4, 5\pi/4, 7\pi/4$ that ${\bf L}$ makes with the $\hat{x}-$axis. The 4 classical loop current states are eigenvalues of operators ${\bf L}_i=(L_{i,x},L_{i,y})$ defined at the unit-cell $i$. We may define a basis, choosing $|{\bf L}_i|$ to be unity,
\be
\label{Lx,y}
(L_{i,x}+iL_{i,y})\ket{\theta}_i=e^{i\theta} \ket{\theta}_i
\ee

The classical statistical mechanics of the Loop-Current state may be derived from the
Ashkin-Teller model, which is given in terms of a pair of Ising spin per unit-cell $\sigma^z_{i}, \tau^z_{i}$, whose eigenvalues, $\pm 1$ specify the $x$ and $y$ components of the direction of the vector ${\bf L}$. The four loop current states can therefore also be denoted as $\ket{\pm1,\pm1}$. The classical Ashkin-Teller model \cite{baxter} is given by \cite{note},
\be
\label{classAT}
H_{AT} = -\sum_{\langle i,j\rangle}[J_1\sigma^z_i\sigma^z_j+J_2\tau^z_i\tau^z_j+
J_4\sigma^z_i\sigma^z_j\tau^z_i\tau^z_j]
\ee

Quantum fluctuations among the four possible directions of order together with dissipation lead \cite{aji-cmv} to a scale invariant spectrum which leads to the observed Marginal Fermi-liquid properties in the quantum-critical regime in the phase diagram, Fig.  (\ref{phase-diagram}). In this paper, we will derive the effect of the quantum fluctuation in the ordered loop current  phase.
In this phase, the kinetic energy locally flips a loop current state in cell $i$ to one of the other three states, just as the transverse field does between the two states of the  transverse field Ising model. Due to this term, the ground state  is a superposition, in each unit-cell, of all the 4 possible directions of ${\bf L}_i$. The excitations consist of local flips between the configurations which spread out spatially through interactions between neighboring sites and acquire dispersion just like spin waves in the transverse-field Ising model.

\subsection{Symmetries}

As explained in Appendix B, in order to treat the four states on the same footing in the quantum-mechanical model, we must enlarge the representation from that of 2 by 2 matrix space to that of 4 by 4 matrices, i.e. we must consider the problem in the $SU(4)$ representation rather than in  the $SU(2)\times SU(2)$ representation. We introduce the following direct products between Pauli matrices.
\be
S^i=\sigma^i\otimes I,\quad T^i=I\otimes\tau^i,\quad K^{ij}=\sigma^i\otimes\tau^j
\ee
The 15 traceless matrices $S$, $T$ and $K$ are also the generators of $SU(4)$ algebra. Their commutators are easy to compute.
\be
&&[S^i,S^j]=2i\epsilon_{ijk}S^k,\quad [T^i,T^j]=2i\epsilon_{ijk}T^k,\\
&&[S^i,T^j]=0,\\
&&[S^i,K^{jk}]=2i\epsilon_{ijl}K^{lk},
\quad[T^i,K^{jk}]=2i\epsilon_{ikl}K^{jl},\\
&&[K^{ij},K^{kl}]=2i\epsilon_{ikm}S^m\delta_{jl}+2i\epsilon_{jln}T^n\delta_{ik}
\ee
Under this choice of basis, the 4 states are labeled by the eigenvalues of $\sigma^z$ and $\tau^z$. We will use the following short-hand notation
\be
\ket{1,1}=\ket{1}_{\sigma}\otimes\ket{1}_{\tau}={1\choose0}\otimes{1\choose0}
=(1,0,0,0)^T.
\ee
And similarly for the other 3 states $\ket{-1,1}, \ket{1,-1}, \ket{-1,-1}$ . In this basis, the classical AT model can be rewritten as
\be
H=-\sum_{\langle i,j\rangle}[J_1 S^3_i S^3_j+J_2T^3_i T^3_j+J_4 K^{33}_i K^{33}_j]
\ee
We now list both the internal and the  lattice symmetries of this classical model.

\subsubsection{Internal symmetries:}

The classical AT model has an obvious global continuous symmetry $U(1)_{S^3}\times U(1)_{T^3}\times U(1)_{K^{33}}$ which is the rotation around $S^3$, $T^3$ and $K^{33}$ generated by the rotation matrix
$U(\theta)=e^{i\theta_1S^3/2}e^{i\theta_2T^3/2}e^{i\theta_3K^{33}/2}$. This is the only continuous symmetry of the model and implies that the phase differences between the four states are arbitrary. Any quantum term will break this symmetry.

The classical model also possess several discrete symmetries. We will argue that the quantum terms should preserve the discrete symmetries.

Like the Ising model, the classical AT model has a global $Z_2$ symmetry which maps $\sigma^z$ to $-\sigma^z$. Since all the interaction terms involve two $\sigma^z$, they stay invariant. Similarly, we also have anther $Z_2$ symmetry which maps $\tau^z$ to $-\tau^z$.

We are more interested in the symmetric case, $J_1=J_2$.
In this case, there is also a discrete symmetry $Z_4$ which is the symmetry under cyclic permutation among the 4 states. Since the coefficient of $S^3$ and $T^3$ terms are the same, the Hamiltonian is the invariant under this transformation. There is another $Z_2$ symmetry which switch $\sigma$ and $\tau$.

\subsubsection{Lattice symmetry}

The 4 states represent the loop current pattern on the square lattice of copper and oxygens. The point group of a square lattice is $D_4$. It consists of 4-fold rotations and reflection about the x-axes, the y-axes and the two diagonal axes. The loop current states can be thought as currents flow along both x and y axes and the values of $\sigma^z$ and $\tau^z$ label the directions of these currents. From these, one can find out how the point group of lattice act on these 4 state.

The 4-fold rotation makes the following transformation
$\ket{1,1}\to\ket{-1,1}\to\ket{-1,-1}\to\ket{1,-1}\to\ket{1,1}$
which is the same as the internal $Z_4$ symmetry. The transformation matrix is given by
\be
C_4=\left(\begin{array}{cccc}
0&0&1&0\\
1&0&0&0\\
0&0&0&1\\
0&1&0&0
\end{array}
\right)
\ee
If we think of $\ket{1,1}$ as a plane vector $(1,1)$, then the above transformation makes a $\pi/2$ anti-clockwise rotation, which realizes the cyclic permutation among the 4 vectors. Then one can find out the following transformation
\be
C_4S^3C_4^{\dagger}=-T^3,\qquad C_4T^3C_4^{\dagger}=S^3,\qquad
C_4K^{33}C_4^{\dagger}=-K^{33}
\ee
Clearly, the classical AT model is invariant under this transformation.

Now we turn to the reflections. The reflection about x-axes switches $\ket{\pm1,1}$ with $\ket{\pm1,-1}$, which is also equivalent to the Ising like internal $Z_2$ symmetry. This reflection can be generated by operator $\sigma(0)=T^1$  Similarly, the reflection about y-axes switches $\ket{1,\pm1}$ with $\ket{-1,\pm1}$ and is generated by operator $\sigma(\pi/2)=S^1$. The reflection about $y=x$ switches $\ket{1,-1}$ and $\ket{-1,1}$ and keep the other two states the same. This is the same as the internal $Z_2$ symmetry $\sigma\leftrightarrow\tau$ and is generated by
\be
\tsigma(\pi/4)=\left(\begin{array}{cccc}
1&0&0&0\\
0&0&1&0\\
0&1&0&0\\
0&0&0&1
\end{array}\right)
\ee
Similarly, the reflection about $y=-x$ switches $\ket{1,1}$ and $\ket{-1,-1}$ and keeps the other two states the same. It is generated by $\tsigma(3\pi/4)=S^1T^1\sigma(\pi/4)$ which is a combination of the 3 internal symmetries. One can see that all the internal discrete symmetries are coincident with the lattice symmetries. We expect that the quantum term preserve all the lattice symmetries, so it also preserve all the internal discrete symmetries.

\subsubsection{Quantum terms allowed.}

Now we want to identify possible terms which lead to the quantum fluctuations among the 4 classical states. In \cite{aji-cmv}, the AT model has been mapped to XY model in the quantum-critical fluctuation regime of the phase diagram. In the loop ordered state, quantum term of AT model can be thought as a discrete version of the quantum term of XY model which makes clockwise and anti-clockwise rotations. As one can see, the unitary operator to generate anti-clockwise rotations is just $C_4$ we introduced. But since det$C_4=-1$, this operator is not a proper rotation in the complex 4D space but also includes a parity transformation e.g. $\ket{1,1}\to-\ket{1,1}$ and keep other 3 states which cannot be realized as a continuous transformation.
Since the quantum evolution of states is obtained by integrating infinitesimal rotations, we expect the anti-clockwise rotation operator $U$ to be a proper rotation, or det$U=1$. (This matter is further discussed in Appendix A). Therefore, we introduce a phase factor to correct the sign of the determinant.
\be
U=\left(\begin{array}{cccc}
0&0&1&0\\
1&0&0&0\\
0&0&0&1\\
0&1&0&0
\end{array}
\right)e^{-i\pi/4}.
\ee
Then the possible quantum terms should be all the possible Hermitian combinations of this unitary operator:
\be
&&U+U^{\dagger}=\frac{1}{\sqrt{2}}[S^1+T^1+(K^{23}-K^{32})]\\
&&i(U-U^{\dagger})=\frac{1}{\sqrt{2}}[S^1+T^1-(K^{23}-K^{32})]\\
&&\frac i2[U^2-(U^{\dagger})^2]=K^{11}
\ee
Since $U^4=-1$, we have $U^2+(U^{\dagger})^2=0$. Therefore the above are the only possible independent Hermitian combinations to quadratic order in $U$. In the most general case, all these terms should be present in the quantum AT model. For our problem, we need to constrain this further to be consistent with the lattice symmetry. The 4-fold rotation operator $C_4$ commutes with $U$, thus there is no constraint due to 4-fold rotation. As discussed, the reflection operators $\tsigma(0)$ and $\tsigma(\pi/2)$ are equivalent to $T^1$ and $S^1$, respectively. And, for the reflection about x and y axes and diagonal directions, we have
\be
&&S^1(U+U^{\dagger})S^1=i(U-U^{\dagger})\\
&&T^1(U+U^{\dagger})T^1=i(U-U^{\dagger}).
\ee
For reflection operator $\sigma(\pm \pi/4)$, we note that
\be
&&\tsigma(\pi/4)(U+U^{\dagger})\tsigma(\pi/4)=i(U-U^{\dagger})\\
&&\tsigma(3\pi/4)(U+U^{\dagger})\tsigma(3\pi/4)=i(U-U^{\dagger}).
\ee
$K^{11}$ is invariant under all these 4 reflections. Once can see that $S^1+T^1$ is even under the reflections and $K^{23}-K^{32}$ is odd. Therefore, the lattice symmetry require the coefficient in front of $U+U^{\dagger}$ and $i(U-U^{\dagger})$ to be the same in order to cancel out the odd part. In conclusion, the proper quantum term that respect all the lattice symmetry should be $S^1+T^1$ and $K^{11}$.

Even if we do not start with the rotation operator $U$, we will get the same results for the quantum term just by imposing  lattice symmetry. In the chosen basis, the only diagonal matrices are $S^3$, $T^3$ and $K^{33}$ and they form a Cartan sub-algebra \cite{LieGroup} of the $SU(4)$ Lie algebra, which is the maximal commuting sub-algebra one can find. All the other 12 generators contains non-zero off-diagonal elements and act as raising and lowering operators, which mix among the 4 states. The most general quantum term should be the superposition of all these 12 generators. Lattice symmetry requires that the quantum term  commute with $S^1$, $T^1$, $K^{11}$ and $\tsigma(\pi/4)$. It is easy to see that $S^1$, $T^1$, $K^{11}$ commute with each other thus they also form a Cartan sub-algebra.  Therefore, the only generators which commute with $S^1$, $T^1$, $K^{11}$ are themselves. $\tsigma(\pi/4)$ switches $S^1$ with $T^1$, thus require the coefficients of $S^1$ and $T^1$ to be the same. Thus the quantum term $S^1+T^1$ and $K^{11}$ is unique up to some unitary transformations.

It may seem strange that only the generators with $1$ component are involved in the quantum term. Actually, this is only due to the specific choice of representation we have made. We can redefine $\tsigma(0)$ and $\tsigma(\pi/2)$ as $\tsigma(0)\ket{1,1}=e^{\phi_1}\ket{1,-1}$ and $\tsigma(\pi/2)\ket{1,1}=e^{\phi_2}\ket{-1,1}$
\be
\tsigma(0)=I\otimes\left(\begin{array}{cc}
0&e^{\phi_1}\\
e^{-\phi_1}&0
\end{array}
\right)=T^1\cos\phi_1-T^2\sin\phi_1\\
\tsigma(\pi/2)=\left(\begin{array}{cc}
0&e^{\phi_2}\\
e^{-\phi_2}&0
\end{array}
\right)\otimes I=S^1\cos\phi_2-S^2\sin\phi_2.
\ee
Since the $D_4$ group can be generated by 2 elements, once the phase factor in $\tsigma(0)$ and $\tsigma(\pi/2)$ are fixed, phase of all other $D_4$ operators are also determined. In general, there are three arbitrary phase difference one can choose, but in the symmetric case, there are only two left. The other symmetry operators are obtained  from $C_4^2=\tsigma(\pi/2)\tsigma(0)$, $\tsigma(\pi/4)=C_4\tsigma(0)$ and $\tsigma(3\pi/2)=C_4^2\tsigma(\pi/4)$. As before, $\tsigma(0)$, $\tsigma(\pi/2)$ and $\tsigma(\pi/2)\tsigma(0)$ form maximal commuting sub-algebra. Therefore the quantum terms are given by $\tsigma(0)+\tsigma(\pi/2)$ and $\tsigma(\pi/2)\tsigma(0)$ with both $x$ and $y$ component present. The new $\tsigma(0)$, $\tsigma(\pi/2)$ are related to the old ones by unitary transformations
\be
\label{gauge-transform}
e^{i(\phi_1T^3+\phi_2S^3)/2}S^1e^{-i(\phi_1T^3+\phi_2S^3)/2}=\tsigma(\pi/2)\nonumber\\
e^{i(\phi_1T^3+\phi_2S^3)/2}T^1e^{-i(\phi_1T^3+\phi_2S^3)/2}=\tsigma(0).
\ee
Thus the results for energy of the collective modes (but not their eigenvectors or the ground state wave-function) should remain the same as in the original gauge.

\section{Dispersion of Collective Modes }
\label{cl}

In the $SU(4)$ formalism, the classical Ashkin Teller model can be written as
\be
H_{AT}=-\sum_{\langle i,j\rangle}[J_1 S^3_i S^3_j+J_2T^3_i T^3_j+J_4 K^{33}_i K^{33}_j]
\ee
According to the discussion of section \ref{sym}, the quantum fluctuations can be described by the Ashkin Teller model with a symmetric quantum terms or transverse field as $S^1+T^1$ and $K^{11}$. The Hamiltonian for the quantum Ashkin-Teller model is therefore given by
\be
H^Q_{AT}&=&\sum_{i}[t(S^1_i+T^1_i)+t'K^{11}_i]\nonumber\\
&-&\sum_{\langle i,j\rangle}[J_1 S^3_i S^3_j+J_2T^3_i T^3_j+J_4 K^{33}_i K^{33}_j]
\ee
As discussed earlier, the basis of  4 states at each lattice site correspond to the 4 types of loop current states. In the AT model, these 4 states are labeled by the eigenvalues of $\sigma^z$ and $\tau^z$ which can be denoted as $|\pm1,\pm1\rangle$. The kinetic term $S^1$ flips the sign of first index and $T^1$ flips the second. Then $K^{11}$ flips both indices together. If we think of the two indices as $x$ and $y$ components of a plane vectors, the 4 states can also be represented by 4 plane vectors. Then $S^1$ and $T^1$ rotates the vector by $\pi/2$ and $K^{11}$ rotates the vectors by $\pi$.  In principle the parameters in this model should be determined from microscopic models. We will treat them as phenomenological parameters and determine them by fitting the observed dispersion of the collective modes, and judge if they have the scale of values expected, which is of the order of  the pseudo-gap energy.

\subsection{The Ground State}
In order to calculate the collective mode dispersion, we first  determine the ground state in the ordered phase at the mean field level. For the loop current ordered phase the classical AT model has ferromagnetic interactions $J_{1,2}>0$ and $J_4<|J_{1,2}|$, the latter to ensure that there is no divergence of the specific heat at the transition. In this phase, we have $\langle S^3\rangle\neq0$, $\langle T^3\rangle\neq0$ and $\langle K^{33}\rangle\neq0$.  The most general uniform quantum ground state is a product over all sites of the  linear combination of the 4 basis states. Since the overall complex factor is redundant, one can parameterize this state by 6 real parameters. Thus we assume the ground state to be
\be
|G\rangle_0&=&\prod_i\cos\frac{\theta_1}{2}\Big(\cos\frac{\theta_2}{2}\ket{1,1}_i
+\sin\frac{\theta_2}{2}\,e^{i\phi_2}\ket{1,-1}_i\Big)\nonumber\\
& &+\sin\frac{\theta_1}{2}\,e^{i\phi_1}\Big(\cos\frac{\theta_3}{2}\ket{-1,1}_i\nonumber\\
& &+\sin\frac{\theta_3}{2}\,e^{i\phi_3}\ket{-1,-1}_i\Big)
\label{gg}
\ee
It is straightforward to compute the energy per site $E_g$ but the result is very complicated and we have to use numerics to minimize $E_g$ to find out parameters $\theta_i$ and $\phi_i$. Since the Hamiltonian does not involve the $2$ component matrices and only $2$ component matrices are imaginary, we expect the 3 phase parameters $\phi_i=0$  for $i=1,2,3$. Thus all the 4 coefficients are real numbers. This can be checked by numerical minimization of the energy. In comparing with experiments, we are more interested in a special case that $J_1=J_2=J$ and the Hamiltonian is invariant under the interchanging between $\sigma$ and $\tau$. In this case, we still have to numerically minimize $E_g$, but we find that the result can be well approximated by a direct product state
\be
|G\rangle&=&\prod_i\left(\cos\frac{\theta}{2}|1\rangle_i
+\sin\frac{\theta}{2}|-1\rangle_i\right)_{\sigma}\nonumber\\
& &\otimes\left(\cos\frac{\theta}{2}|1\rangle_i
+\sin\frac{\theta}{2}|-1\rangle_i\right)_{\tau}
\ee
which is parameterized by one angle $\theta$. Then the angle is determined by equation,
\be
t+t'\sin\theta+4J\sin\theta+4J_4\sin\theta\cos^2\theta=0.
\ee
We will use this direct product ground state $|G\rangle$ for latter discussions.
The energy of this direct product ground state is only $3\%$ higher than the most general ground state, so it is a good approximation.
We emphasize that this wave-function is in a particular gauge; the change of gauge,  Eqs. (\ref{gauge-transform}), changes the ground state wave-function.

\subsection{Collective Modes}
To compute the spin wave like collective modes in this quantum Ashkin-Teller model, we generalize the Holstein-Primakoff transformation for $SU(2)$  to $SU(4)$. Introducing three boson operators $b_i$, $c_i$ and $d_i$, we have the following boson operator representations for the Ashkin-Teller model (For ease of notation, we omit  the lattice site label $i$ ):
\be
&&S^1=U c+\cd U+\bd d+\dd b,\nonumber\\
&&S^3=1-2\cd c-2\dd d,\nonumber\\
&&T^1=U b+\bd U+\cd d+\dd c,\nonumber\\
&&T^3=1-2\bd b-2\dd d,\nonumber\\
&&K^{11}=U d+\dd U+\bd c+\cd b,\nonumber\\ 
&&K^{33}=1-2\bd b-2\cd c,\nonumber\\
&&K^{13}=U c+\cd U-\bd d-\dd b,\nonumber\\
&&K^{31}=U b+\bd U-\cd d-\dd c,\nonumber
\ee
with $U=(1-\bd b-\cd c-\dd d)^{1/2}$. In this representation, we take the {\it classical} ground state such that $\langle S^3_i\rangle=\langle T^3_i\rangle=\langle K^{33}_i\rangle=1$. Thus the {\it classical} ground state is just $|1,1\rangle$  and $\bd_i,\,\cd_i,\,\dd_i$ are the creating operators of states $|1,-1\rangle$, $|-1,1\rangle$ and $|-1,-1\rangle$ at site $i$ from the classical  ground state respectively.

Due to the quantum terms, the mean field ground state $|G\rangle$ is a superposition of the 4 states at each site. In order to make use of the Holstein-Primakoff transformation, we can make a basis rotation to transform the ground state $\ket{G}$ to $\prod_i|1,1\rangle_i$ in the new basis.

In the following calculations, we still keep $J_1$ and $J_2$ as two different parameters. In the last step, we will take $J_1=J_2=J$. For the direct product ground state $|G\rangle=\prod_i\left(\cos\frac{\theta_1}{2}|1\rangle_i
+\sin\frac{\theta_1}{2}|-1\rangle_i\right)_{\sigma}
\otimes\left(\cos\frac{\theta_2}{2}|1\rangle_i
+\sin\frac{\theta_2}{2}|-1\rangle_i\right)_{\tau}$, the rotation we need is just two rotations around $y$ axes by angle $\theta_{1,2}$ in the $\sigma$ and $\tau$ space.
\be
S^3\to S^3\cos\theta_1-S^1\sin\theta_1\quad T^3\to T^3\cos\theta_2-T^1\sin\theta_2\nonumber\\
S^1\to S^3\sin\theta_1+S^1\cos\theta_1\quad T^1\to T^3\sin\theta_2+T^1\cos\theta_2\nonumber
\ee
Then the Hamiltonian is
\be
&&H=\sum_i[t(S_i^3s_1+S_i^1c_1)+t(T_i^3s_2+T_i^1c_2)\nonumber\\
&&\qquad+t'(K^{33}_is_1s_2+K^{31}_is_1c_2+K^{13}_ic_1s_2+K^{11}_ic_1c_2)]\nonumber\\
&&-\sum_{i,j}\Big[J_1(S_i^3S_j^3c^2_1-S_i^3S_j^1c_1s_1-S_i^1S_j^3c_1s_1+S_i^1S_j^1s^2_1)\nonumber\\
&&\qquad+J_2(T_i^3T_j^3c^2_2-T_i^3T_j^1c_2s_2-T_i^1T_j^3c_2s_2+T_i^1T_j^1s^2_2)\Big]\nonumber\\
&&-J_4\sum_{i,j}(K^{33}c_1c_2-K^{31}c_1s_2-K^{13}s_1c_2+K^{11}s_1s_2)_i\nonumber\\
&&\times(K^{33}c_1c_2-K^{31}c_1s_2-K^{13}s_1c_2+K^{11}s_1s_2)_j.
\label{Ht}
\ee
with $c_{1,2}=\cos\theta_{1,2}$ and $s_{1,2}=\sin\theta_{1,2}$. The $\theta_{1,2}$ can be solved from the equations.
\be
&&t+t's_2+4J_1s_1+4J_4s_1c_2^2=0\nonumber\\
&&t+t's_1+4J_2s_2+4J_4s_2c_1^2=0\nonumber
\ee

\subsubsection{Simple Case, $J_4 =t' = 0$}

We can start from the simple case with $J_4=0$ and $t'=0$. In this case, the Hamiltonian is two decoupled Ising models. We can plug in the boson transformation, expand the square root up to quadratic order. Since we only care about the collective mode dispersion, the constant term can be ignored. The linear terms in boson operators will cancel out due to the minimization condition of the ground state energy. Thus we are only left with quadratic terms,
\begin{widetext}
\be
&&H_0=t\sum_i[(-2\cd c-2\dd d)_i s_1+(\bd d+\dd b)_ic_1]
-J_1\sum_{i,j}\Big[(-2\cd c-2\dd d)_i c_1^2+(-2\cd c-2\dd d)_j c_1^2\nonumber\\
&&\quad -(\bd d+\dd b)_i s_1c_1-(\bd d+\dd b)_js_1c_1+(c+\cd)_i(c+\cd)_js_1^2\Big]
+\mbox{terms}\,(b\leftrightarrow c, 1\leftrightarrow 2).
\ee
This equation can be simplified by using identity $-ts_i+4J_ic_i^2=4J_i$. Note that $(\bd d+\dd b)$ and $(\cd d+\dd c)$ terms cancels out, thus operator $d$ decoupled from $b$ and $c$ . In momentum space we have
\be
&&H_0=\sum_{\vk}(4J_1-2J_1s_1^2\fk)(\cd_{\vk}c_{\vk}+c_{-\vk}\cd_{-\vk})
-2J_1s_1^2\fk(c_{\vk}c_{-\vk}+\cd_{\vk}\cd_{-\vk})\nonumber\\
&&+(4J_2-2J_2s_2^2\fk)(\bd_{\vk}b_{\vk}+b_{-\vk}\bd_{-\vk})
-2J_2s_2^2\fk(b_{\vk}b_{-\vk}+\bd_{\vk}\bd_{-\vk})+8(J_1+J_2)\dd_{\vk}d_{\vk}.
\ee
\end{widetext}
One can see the that the three sets of 3 bosons are decoupled. The $b$ and $c$ part are two copy of Ising models and the $d$ part is free boson. Making use of standard Bogoliubov transformation we can diagonalize the above Hamiltonian and find the following dispersions
\be
&&\omega_1=2[(4J_1)^2-t^2\fk]^{1/2},\nonumber\\
&&\omega_2=2[(4J_2)^2-t^2\fk]^{1/2},\nonumber\\
&&\omega_3=8(J_1+J_2).\nonumber
\ee
In addition to the two Ising model modes, we also find a third mode which is dispersion-less.
The two Ising modes are generated by bosons operators $b$ and $c$ which only rotate $\sigma$ or $\tau$ space alone. Since the ground state is a direct product, the excited state generated by modes $b$ and $c$ are also direct products.
\be
\exp(-i\phi b)|1,1\rangle&\approx&|1,1\rangle-i\phi|1,-1\rangle\nonumber\\
&=&|1\rangle\times\Big(|1\rangle-i\phi|-1\rangle\Big).
\ee
On the other hand, the third mode which is generated by boson operator $d$ cannot be written as a direct product. For example,
\be
\exp(-i\phi d)|G_0\rangle\approx|1,1\rangle-i\phi|-1,-1\rangle,
\ee
which cannot be written in the form: $|\psi_{\sigma}\rangle\otimes|\psi_{\tau}\rangle$.

The constant energy feature of the third mode can be explained as follows.
The Hamiltonian with $J_4=0$ in the rotated basis is the first 2 lines of Eq. (\ref{Ht}). If we linearize the Hamiltonian, then we have following terms involving $S$,
\be
-\sum_i 4JS^3-J\sum_{i,j}S^1_iS^1_js_1^2.
\ee
The first terms describes the rotation of S spin under a $z$ direction constant external field for all lattice sites. If we only keep this term, we will only get a constant dispersion $4J$. Since now the ground state is up spin for all lattice sites and $S^1$ can flip up spin to down spin, the term $S^1_iS^1_j\sin^2\theta$ describes an interaction between the flip in site i and neighboring site j. Thus the spin flip can propagate and give a plane wave momentum dependence like $\fk s_1^2$, which agrees with our previous result. The third mode corresponds to spin flip: $|1,1\rangle\to|-1,-1\rangle$. In order to make this flip propagate, one need a term like $K^{11}_iK^{11}_j$ in the Hamiltonian. But when $J_4=0$ there is no such terms, thus the third modes is constant when $J_4=0$.
When $J_4\neq0$, we have a term $K^{11}_iK^{11}_js_1^2s_2^2$ in Eq (\ref{Ht}).
and the third mode will acquire momentum dependence.

\subsubsection{General Case}

Now we turn to the general case when $J_4\neq0$ and $t'\neq0$. The $t'$ and $J_4$ terms involve operators like $K_{11}$ and $K_{33}$ which couple the $\sigma$ and $\tau$ spin and split the two degenerate modes we obtained in the decoupled case. We denote these two terms by $H'$. Transforming to the boson representation we have
\begin{widetext}
\be
&&H'=t'\sum_i\Big[(-2\bd b-2\cd c)_is_1s_2+(-\cd d-\dd c)_is_1c_2
+(-\bd d-\dd b)_ic_1s_2+(\bd c+\cd b)_ic_1c_2\Big]\nonumber\\
&&-2J_4\sum_i\Big[2(-2\bd b-2\cd c)_i c_1^2c_2^2+2(\cd d+\dd c)_ic_1^2s_2c_2
+2(\bd d+\dd b)_is_1c_1c_2^2+2(\bd c+\cd b)_is_1c_1s_2c_2\nonumber\\
&&+\sum_j(b+\bd)_i(b+\bd)_jc_1^2s_2^2+\sum_j(c+\cd)_i(c+\cd)_js_1^2c_2^2
+\sum_j(d+\dd)_i(d+\dd)_js_1^2s_2^2\nonumber\\
&&+2\sum_j(b+\bd)_i(c+\cd)_js_1c_1s_2c_1-2\sum_j(b+\bd)_i(d+\dd)_js_1c_1s_2^2
-2\sum_j(c+\cd)_i(d+\dd)_js_1^2s_2c_2\Big]. \nonumber
\ee
We can make use of identities like $-ts_1-t's_1s_2+4J_1c_1^2+4J_4c_1^2c_2^2=4(J_1+J_4c_2^2)$ to simplify the above equation. Transforming to momentum space and combining with $H_0$, the total Hamiltonian is (we drop the subscript $\vk$ to simplify the notation)
\be
&&H=\sum_{\vk}(J_2+J_4c_1^2)[8\bd b-2s_2^2\fk(b+\bd)^2]
+(J_1+J_4c_2^2)[8\cd c-2s_1^2\fk(c+\cd)^2]\nonumber\\
&&\qquad\qquad +\Big[8(J_1c_1^2+J_2c_2^2)-2t(s_1+s_2)\Big]\dd d-2J_4s_1^2s_2^2\fk(d+\dd)^2\nonumber\\
&&+2(t+4J_1s_1)c_1(\bd d+\dd b)+2(t+4J_2s_2)c_2(\cd d+\dd c)+(t'-4J_4s_1s_2)c_1c_2(\bd c+\cd b)\nonumber\\
&&-4J_4\Big[s_1c_1s_2c_1\fk(b+\bd)(c+\cd)
-s_1c_1s_2^2\fk(b+\bd)(d+\dd)-s_1^2s_2c_2\fk(c+\cd)(d+\dd)\Big],
\label{HP}
\ee
\end{widetext}
with $\fk=(\cos k_x+\cos k_y)/2$.
We can rewrite the Hamiltonian (\ref{HP}) in a matrix form $H=\psi^{\dagger}M\psi$
with $\psi=(\bd,b,\cd,c,\dd,d)^T$ and the 6 by 6 symmetric $M$. There are only 12 independent elements because the matrix elements are the same under the interchange $1\leftrightarrow2$, $3\leftrightarrow4$ and $5\leftrightarrow6$ for both indices or by interchanging the position of the two indices $ij\leftrightarrow ji$. These matrix elements are easy to read off from Eq. (\ref{HP}).

Since the Hamiltonian is quadratic, one can use standard Bogoliubov transformation to diagonalize the above Hamiltonian and find the collective mode dispersions. We introduce the following quasi-particle creation and annihilation operators
\be
b=u_{11}\alpha+v_{11}\alpha^{\dagger}+u_{12}\beta+v_{12}\beta^{\dagger}+u_{13}\gamma+v_{13}\gamma^{\dagger}\nonumber\\
c=u_{21}\alpha+v_{21}\alpha^{\dagger}+u_{22}\beta+v_{22}\beta^{\dagger}+u_{23}\gamma+v_{23}\gamma^{\dagger}\nonumber\\
d=u_{31}\alpha+v_{31}\alpha^{\dagger}+u_{32}\beta+v_{32}\beta^{\dagger}+u_{33}\gamma+v_{33}\gamma^{\dagger}.\nonumber
\ee
This introduces 18 real parameters $u_{ij},v_{ij}$ for $i,j=1,2,3$. In order to make the quasi-particles to be bosons we require that the only non zero commutators are $[\alpha,\alpha^{\dagger}]=[\beta,\beta^{\dagger}]=[\gamma,\gamma^{\dagger}]=1$. This leads to the following constraint
\be
&&\sum_{m=1}^3(u_{im}^2-v_{im}^2)=1\nonumber\\
&&\sum_{m=1}^3\epsilon_{ijk}(u_{im}v_{jm}-v_{jm}u_{im})=0\nonumber\\
&&\sum_{m=1}^3\epsilon_{ijk}(u_{im}u_{jm}-v_{im}v_{jm})=0.
\label{cons}
\ee
for $i,j,k=1,2,3$. There are 9 constraints, thus we are left with 9 free parameters.

Thus there are only 9 independent  elements in Hamiltonian so that the off-diagonal 9 can be set to zero by adjusting the 9 free parameters among $u$ and $v$. Then we are left with a diagonal Hamiltonian and 3 dispersion modes. But it is very difficult to explicitly solve the 18 real parameters $u$ and $v$. It is easier to determine the the dispersions directly. But this can not be done by directly diagonalize $M$, because $u$ and $v$ do not form a unitary matrix. Introducing $\phi=(\alpha^{\dagger},\alpha,\beta^{\dagger},\beta,\gamma^{\dagger},\gamma,)^T$, the Bogoliubov transformation can be rewritten as
$\psi=U\phi$. One can verify that $U^{\dagger}U\neq1$. This is because $\psi^{\dagger}\psi\neq\phi^{\dagger}\phi$ and $\psi^{\dagger}\psi$ cannot be treat as the norm of the vector $\psi$. This can be fixed by introducing matrix $\sigma=\mbox{diag}\{1,-1,1,-1,1,-1\}$, then we have $\psi^{\dagger}\sigma\psi=[b,\bd]+[c,\cd]+[d,\dd]=3$ and similarly $\phi^{\dagger}\sigma\phi=3$. Thus we have $U^{\dagger}\sigma U=\sigma$. This can also be written as
$\sigma U^{\dagger}\cdot\sigma U=I$ and $\sigma U\cdot\sigma U^{\dagger}=I$. One can verify that the last equation is consistent with the constraints in Eq. (\ref{cons}).

Suppose we have determined the parameters $u$ and $v$ that diagonalize the Hamiltonian, then we have
$U^{\dagger}MU=\mbox{diag}\{\omega_1,\omega_1,\omega_2,\omega_2,\omega_3,\omega_3\}$ and $\omega_{1,2,3}$ are the three excitation modes. We can rewrite this as
\be
\sigma U^{\dagger}\cdot M\sigma\cdot\sigma U=\mbox{diag}\{\omega_1,-\omega_1,\omega_2,-\omega_2,\omega_3,-\omega_3\}
\ee
It is easy to see that
\be
&&\mbox{det}(\sigma U^{\dagger}\cdot M\sigma\cdot\sigma U-\omega I)\nonumber\\
&&=\mbox{det}(\sigma U^{\dagger})\mbox{det}(M\sigma-\omega I)\mbox{det}(\sigma U)=\mbox{det}(M\sigma-\omega I)\nonumber
\ee
thus matrix $\sigma U^{\dagger}\cdot M\sigma\cdot\sigma U$ and $M\sigma$ have same eigenvalues. We can determine the dispersion of the three modes by solving the equation
\be
\mbox{det}(M\sigma-\omega I)=0
\ee
which is a cubic equation in $\omega^2$. We determine the roots numerically. We again find 3 collective modes but now all of them are weakly dispersive.

Actually we are more interested in the special case $J_1=J_2$. In this case, one can further simplify the equation by introducing symmetrized and anti-symmetrized variables $a_1=(b+c)/\sqrt{2}$ and $a_2=(b-c)/\sqrt{2}$. Then the 6 by 6 matrix will decompose into one 4 by 4 matrix for symmetrized operator $a_1$ and $d$ and one 2 by 2 matrix for anti-symmetrized operator $a_2$. Now the eigenvalues of symmetrized operator $a_1$ and $d$ are
\begin{widetext}
\be
&&\omega_{1,3}=\bigg[2(A_sA_d+B_sB_d+2C_sC_d)\pm2\sqrt{(A_sA_d+B_sB_d+2C_sC_d)^2
-4(A_sB_s-C_s^2)(A_dB_d-C_d^2)}\bigg]^{1/2}
\ee
with $A_{s,d}=A_1\pm A_2$ and the same for $B_{s,d}$ and $C_{s,d}$. Here we have $A_1=2J(2-s^2\fk)-2J_4c^2[-2+(1+2\fk)s^2]+t'c^2/2$, $A_2=-2(J+2J_4c^2)s^2\fk$, $B_1=8Jc^2-2ts-2J_4s^4\fk$, $B_2=-2J_4s^4\fk$, $C_1=\sqrt{2}c(t+4Js+2J_4s^3\fk)$ and $C_2=2\sqrt{2}J_4s^3c\fk$ with $s=\sin\theta$ and $c=\cos\theta$.
The eigenvalues of anti-symmetrized operator $a_2$ is
\be
\omega_2=\bigg[\Big(8J+2J_4c^2(2+s^2)-t'c^2\Big)
\Big(8J+2J_4c^2(2+s^2)-t'c^2-8Js^2\fk\Big)\bigg]^{1/2}\label{om2}
\ee
\end{widetext}

In order to get a qualitative picture of these three modes, we can numerically compute some typical eigenvectors. For example, take the point at zone boundary $k_x=\pi/a,\,k_y=0$. The three eigen-modes are created by the following quasi-particle operators
\be
&&\alpha^{\dagger}=0.67(\bd+\cd)-0.3\dd,\nonumber\\
&&\beta^{\dagger}=0.71(\bd-\cd),\nonumber\\
&&\gamma^{\dagger}=0.21(\bd+\cd)+0.95\dd\nonumber
\ee
One can see that the first and third modes symmetrically mix  $|1,1\rangle\to|1,-1\rangle+|-1,1\rangle$ and  $|1,1\rangle\to|-1,-1\rangle$ . The first mode has more weight on the former and third one has weight on the latter.  On the other hand the second mode mixes $|1,1\rangle\to|1,-1\rangle-|-1,1\rangle$. anti-symmetrically.

In the decoupled limit, we have two degenerate Ising modes for bosons $b$ and $c$. In other word, these two modes are propagation of 90 degree flip. The third mode is constant corresponding to boson $d$ or 180 degree flip. In general case with non-zero $J_4$ and $t'$, there is no degeneracy and the third mode is also dispersive. In order to understand the qualitative effects of $J_4$, we first consider that $t'=0$ and $J_4$ is small and then make a perturbation expansion as follows
\begin{widetext}
\be
&&\omega_1=\sqrt{16J^2-\fk t^2}+\frac{16J^2-t^2}{\sqrt{16J^2-\fk t^2}}
\left[1-(1+\fk)\frac{t^2}{32J^2}-\fk\frac{t^4}{1024J^4}\right]\frac{J_4}{J}\nonumber\\
&&\omega_2=\sqrt{16J^2-\fk t^2}+\frac{32J^2-\fk t^2}{\sqrt{16J^2-\fk t^2}}
\left[\frac12-\frac{3t^2}{64J^2}+\frac{t^4}{1024J^4}\right]\frac{J_4}{J}\nonumber\\
&&\omega_3=8J+\left[\frac{t^2}{2J^2}-(4+\fk)\frac{t^4}{128J^4}\right]J_4\nonumber
\ee
We can also assume that $J_4=0$ and $t'$ is small then make a perturbation expansion as follows
\be
&&\omega_1=\sqrt{16J^2-\fk t^2}+\frac{J^2}{\sqrt{16J^2-\fk t^2}}
\left[-2+(2+5\fk)\frac{t^2}{16J^2}-\fk\frac{t^4}{256J^4}\right]\frac{t'}{J}\nonumber\\
&&\omega_2=\sqrt{16J^2-\fk t^2}+\frac{J^2}{\sqrt{16J^2-\fk t^2}}
\left[2+(-2+3\fk)\frac{t^2}{16J^2}+\fk\frac{t^4}{256J^4}\right]\frac{t'}{J}\nonumber\\
&&\omega_3=8J+\frac{t^2}{8J^2}t'\nonumber
\ee
\end{widetext}
Qualitatively, both $J_4$ and $t'$ split the dispersions of $\omega_1$ and $\omega_2$, but $J_4$ also changes the relative dispersion of these two modes. More precisely, positive $J_4$ makes $\omega_1$ more dispersive than $\omega_2$ and negative $J_4$ makes $\omega_1$ less dispersive than $\omega_2$. On the other hand, $t'$ mostly shifts the whole curve and makes both the modes more dispersive. Positive $t'$ pushes $\omega_1$ up and push $\omega_2$ down, while negative $t'$ does the opposite.
\begin{figure}
\centerline{\includegraphics[width=0.5\textwidth]{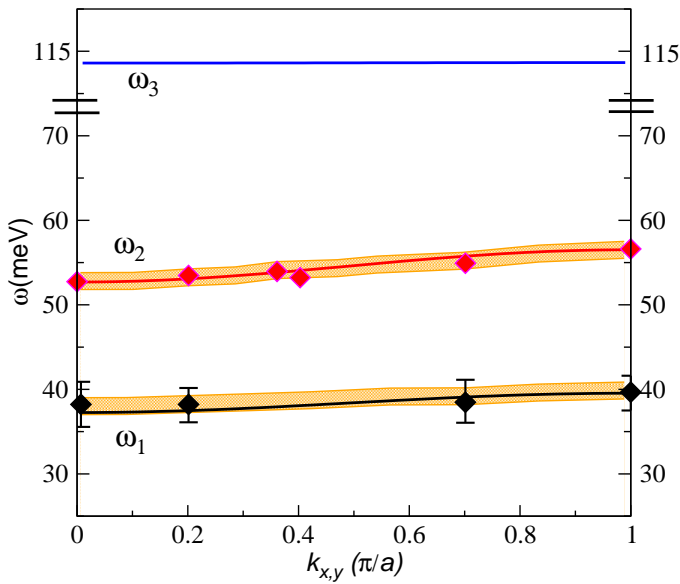}}
\caption{The collective mode dispersions of Ashkin-Teller model with negative $J_4$ as a function of $k_x$ along the $k_x=k_y$ direction. The black curve stands for $\omega_1$ and red curve stands for $\omega_2$. The parameters are taken to fit the experimental results in $Hg1201, T_c=65K$. The blue curve gives the calculated third mode.}
\label{dispersion}
\end{figure}

\subsection{Damping}

The principal source of the damping of the collective modes is the decay into particle-hole pairs, of similar energy and same symmetry as the collective modes, provided by $H_{F-coll}$. There exists a pseudo-gap in the single-particle fermion spectra in the same region of the phase diagram as the observed order. Therefore for energies of up to about 50 meV in the well under-doped region, damping of the collective modes is expected not to obscure their visibility. Still, it is surprising that the two observed branches of modes \cite{coll-modes} in $Hg1201, T_c = 61 K$ have a damping smaller than the experimental resolution of about 5 meV. This may be due to the large magnitude of the order parameter so that most of the orbital current spectral weight of the appropriate symmetry which in the non-interacting model is in the incoherent fermion excitations is transferred to the collective excitations. It is possible that the third branch, which is at about 100 meV has a significantly higher damping.

\section{Comparison with measured Dispersion}

We Choose parameters to reproduce the experiments \cite{coll-modes} in $Hg1201$ with $T_c = 61 K$. The results are shown in Fig. (\ref{dispersion}). In the experiments, one of the collective modes is at $40\pm 5$ meV another one at $50 \pm 5$ meV at $k=0$. The dispersion across the Brillouin zone in both the $(11)$ and the $(10)$ directions for the higher energy mode is $5 \pm 5$ meV, while the lower mode is even less dispersive. We also have the very important constraint from the thermodynamics of the AT model \cite{baxter, sudbo} that to have a transition with no divergence in the specific heat  $-1 < J_4/J < 0, J>0$. In Fig. (\ref{dispersion}), the following parameters are used:
\be
&&t=3.9\,\mbox{meV},\quad J=7.25\,\mbox{meV},\nonumber\\
&&J_4=-0.2J,\quad t'=-2.15t.
\ee
These give that $\sin\theta \approx -0.257$. The dispersion width of the first mode is about $10\%$ of the energy gap and second one is around $5\%$ of the energy gap. The results are shown in figure \ref{dispersion}. These dispersions  change with doping;  we use a doping where a lot of data is available.

As discussed in Appendix D, the highest energy branch with eigenvalue  $\omega_3$, is primarily an excitation with angular momentum 2 with a small admixture of angular momentum 1. The former are undetectable in neutron scattering experiments. This branch also has a higher energy than has currently been addressed by inelastic neutrons; it may also be an over-damped excitations since it lies well above the pseudo-gap energy in the single-particle spectra.

\section{conclusions}

We find that with reasonable parameters, one can fit the measured dispersion of the rather unique collective modes in the loop ordered phase.
The experimental discovery of these collective modes, whose appearance as a function of temperature coincides with the transition temperature of the loop ordered phase and whose intensity is consistent in its variation as the square of the measured order parameter as a function of temperature, adds further confidence to the existence of loop order. Indeed, there is no calculation we know of in which other proposed form of order give multiple weakly dispersive collective modes.

The energy of all the modes calculated should $\to 0$ for momentum ${\bf q} \to 0$, as $T \to T^*(x)$, the loop order temperature. We suggest experiments to verify this. It also follows that the spectral weight of these modes comes at the expense of the unusual local quantum fluctuations \cite{aji-cmv} in the marginal fermi-liquid region of the phase diagram of the cuprates. Accordingly, we expect a strong diminution of the low energy part of such fluctuations in the pseudo-gap region of the phase diagram.

When quantum terms are included, the ground state is a product over all unit-cells of the  linear combinations of the four classical configurations. Beyond mean-field calculations, the ground state is more complicated and includes effects due to the zero point fluctuations of the collective modes. Since the modes are Ising-like,  this is a small change. We have estimated that the quantum correction to the expectations $\langle S^3\rangle$ $\langle T^3\rangle$ and $\langle K^{33}\rangle$ are smaller than $1\%$.

The mean-field ground state discovered here has important bearing on the effective angle with respect to the c-axis, deduced by elastic polarized neutron scattering experiments \cite{loop-order-expt}. This is discussed in an accompanying paper. It also may have consequences for the gap in one-particle fermion spectra in the pseudo-gap regime. We hope to discuss this important matter in the near future.

\acknowledgments
We wish to acknowledge useful discussions of the experimental results and comparison with the calculations with Philippe Bourges, Martin Greven, Yuan Li, Herb Mook and Yvan Sidis. We have also benefitted from discussions with Vivek Aji and Thierry Giamarchi on several important issues.

\appendix
\section{AT model in $SO(6)$ spinor representation}

In this appendix, we will rewrite the Ashkin-Teller model generators in terms of $SO(6)$ spinor representation \cite{LieGroup}. In this formalism, one can explicitly show that  the spinor representation that requires $U^4=-1$ instead of $U^4=1$. From group theory we know that locally $SU(4)$ is equivalent to $SO(6)$. More precisely, $SU(4)$ is a double cover of $SO(6)$. Thus the fundamental representation of $SU(4)$ we have used in the paper corresponds to the spinor representation of $SO(6)$. We can express $U$ in terms of this spinor representation. The minimal dimension of gamma matrices of $SO(6)$ is 8. One can use the eigenvalues of $\gamma_7$ to define chiral spinor which is dimension 4, same as the fundamental representation of $SU(4)$. We can introduce the following 6 gamma matrices
\be
\gamma_a&=&(\sigma^x\otimes S^2,\,\sigma^x\otimes S^3,\,\sigma^x\otimes K^{11},\,\nonumber\\
& &\quad\sigma^x\otimes K^{12},\,\sigma^x\otimes K^{13},\,\sigma^y\otimes I_4)
\ee
and one can verify that they satisfy $\{\gamma_a,\gamma_b\}=2\delta_{ab}$. Then the 15 generators of $SO(6)$ are given by $S_{ab}=\frac{1}{4i}[\gamma_a,\gamma_b]$ and they correspond to the 15 generators of $SU(4)$. They can be explicitly written as
\be
S_{ab}&=&\frac12\Big(I\otimes S^1,\,I\otimes T^{1,2,3},\,I\otimes K^{ij},\,\nonumber\\
& &\quad\sigma^z\otimes S^{2,3},\,\sigma^z\otimes K^{1j}\Big)
\ee
with $i=2,3$ and $j=1,2,3$. Note that all the above generators can be decomposed as two 4 by 4 blocks. The same is true for $\gamma_7=i\gamma_1\cdots\gamma_6=\sigma^z\otimes I_4$. The 8 dimension spinor thus decomposes into two 4 dimension chiral spinors according to $\gamma_7=\pm1$. The rotation operator can be written as
\be
U=\exp\Big[-i\frac{\pi}{4}(S^1+T^1-K^{23}+K^{32}-K^{11})\Big]
\ee
The corresponding operator acting on the spinor space is
\be
U'=\exp\Big[-i\frac{\pi}{2}(S_{12}+S_{45}+S_{25}-S_{14}-S_{36})\Big]
\ee
So this is a $\pi/2$ rotation of a 2d plane inside a 6D space. Then $U'^4$ is a $2\pi$ rotation in the 6D space. When it act on the spinors, it gives an extra minus sign, just as in the 4D Dirac spinor case. The same thing happened to $U^4$ since it act on the chiral spinors.

\section{Need for $SU(4)$ formalism}

In order to treat the 4 states on the equal footing, we enlarge the the $SU(2)\times SU(2)$ group to $SU(4)$ group. Actually, in the classical AT model, only $S^3$, $T^3$ and $K^{33}$ appear, and all of them are diagonal matrices and commute with each other and can be treat as numbers. Thus at the classical level there is no difference between  $SU(2)\times SU(2)$ and $SU(4)$ formalism. But for the quantum model, this is no longer true. In the $SU(2)\times SU(2)$ formalism, all the quantum states are direct product $S^2\times S^2$ and they can be parametrized by 4 real numbers. There are only two sets of raising and lowering operators which can be used as quantum flipping terms. In $SU(4)$ formalism, the space of all possible quantum states is $CP_3$ which is parametrized by 6 real numbers. There 6 sets of raising and lowering operators and they exhaust all possible rotations in $CP_3$.

Therefore $SU(2)\times SU(2)$ can not make the most general unitary evolution in the 4 state system. Thus It can not treat all the rotations among the 4 states on the same footing. In the linear order, the only quantum flipping term is $\sigma^x$ or $\tau^x$ or equivalent terms, which can switching between $\sigma^z=\pm1$ or $\tau^z=\pm1$. Other types of rotation such as cyclic permutation $U$ can not be expressed as linear combinations of $SU(2)\times SU(2)$ generators. They have to be expressed in terms of high order power terms.

\begin{figure}
\centerline{\includegraphics[width=0.4\textwidth]{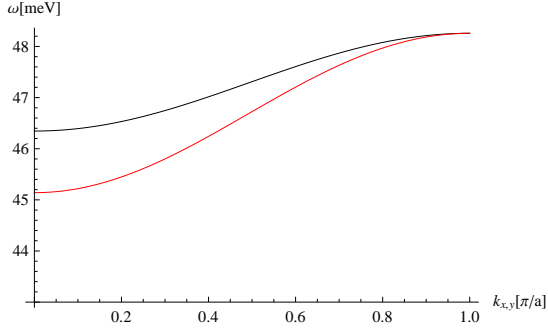}}
\caption{The two modes in the $SU(2)\times SU(2)$ formalism.}
\label{SU2}
\end{figure}

\section{The collective modes in the $SU(2)\times SU(2)$ Formalism}

Nevertheless, It is interesting to compute the collective modes of quantum AT model without enlarging the basis to $SU(4)$ but restrict to $SU(2)\times SU(2)$ group and compare the results. In this way, $\sigma^i$ and $\tau^i$ are just 2 by 2 matrices. They correspond to $S^i$ and $T^i$ matrices in the $SU(4)$ basis. There is no operators correspond to $K^{ij}$ in $SU(2)\times SU(2)$ basis and all the quantum states are direct product of two sets of Ising spin states. Now we only have to introduce two HP bosons thus there are only two modes.

Now we introduce the HP boson as follows.
\be
&&\sigma^x=U_1b+\bd U_1,\quad\sigma^y=-i(U_1b-\bd U_1),\nonumber\\
&&\sigma^z=1-2\bd b\nonumber\\
&&\tau^x=U_2c+\cd U_2,\quad\tau^y=-i(U_2c-\cd U_2),\nonumber\\
&&\tau^z=1-2\cd c\nonumber
\ee
with $U=(1-\bd b)^{1/2}$ and with $U_2=(1-\cd c)^{1/2}$.

We consider the simplest case.
\be
&&H_{QAT}=\sum_{i}[t(\sigma^x_i+\tau^x_i)]\nonumber\\
&&-\sum_{\langle i,j\rangle}[J\sigma^z_i\sigma^z_j+J\tau^z_i\tau^z_j+
J_4\sigma^z_i\sigma^z_j\tau^z_i\tau^z_j]
\ee
Now the ground state is
\be
|G\rangle&=&\left(\cos\frac{\theta}{2}|1\rangle
+\sin\frac{\theta}{2}|-1\rangle\right)_{\sigma}\nonumber\\
& &\otimes\left(\cos\frac{\theta}{2}|1\rangle
+\sin\frac{\theta}{2}|-1\rangle\right)_{\tau}
\ee
with $\theta$ being determined by $t+4Js+4J_4sc^2=0$ with $s=\sin\theta$ and $c=\cos\theta$.

As before, we should first make a basis rotation to make the ground state satisfying $\langle\sigma^z\rangle=\langle\tau^z\rangle=1$. Then we can plug in the HP representation and expand to quadratic order. In the momentum space, we find
\begin{widetext}
\be
H&=&\sum_{\vk}\Big(-2ts\bd_k b_k-2J[-4(\bd_k b_k)c^2+(b_k+\bd_k)^2f_ks^2]
\nonumber\\
& &-2ts\cd_k c_k-2J[-4(\cd_k c_k)c^2+(c_k+\cd_k)^2f_ks^2]\nonumber\\
&-&2J_4[-4c^4(\bd_k b_k+\cd_k c_k)+2s^2c^2(b_k+\bd_k)(c_k+\cd_k)
+(b_k+\bd_k+c_k+\cd_k)^2f_ks^2c^2]\Big)
\ee
Let $\psi=(\bd,b,\cd,c)^T$, then the Hamiltonian can be written as
$H=\psi^{\dagger}M\psi$ with
\be
M=\left(\begin{array}{cccc}
4J'(2-s^2f_k) & -4J's^2f_k & -4J_4s^2c^2(1+f_k) & -4J_4s^2c^2(1+f_k)\\
-4J's^2f_k & 4J'(2-s^2f_k) & -4J_4s^2c^2(1+f_k) & -4J_4s^2c^2(1+f_k)\\
-4J_4s^2c^2(1+f_k) & -4J_4s^2c^2(1+f_k) & 4J'(2-s^2f_k) & -4J's^2f_k\\
-4J_4s^2c^2(1+f_k) & -4J_4s^2c^2(1+f_k) & -4J's^2f_k & 4J'(2-s^2f_k)
\end{array}\right)
\ee
Here $J'=J+J_4c^2$.

Then we can diagonalize the above matrix and find the dispersion of the two collective modes
\be
\omega_k=8\sqrt{(J+J_4c^2)\Big[(J+J_4c^2)(1-s^2f_k)\pm J_4s^2c^2(1+f_k)\Big]}
\ee
\end{widetext}

If we let $J_4=0$, then we get $\omega_k=8J(1-s^2f_k)^{1/2}$ which is the same as the Ising model solution. Since $f_k=-1$ for $k_x=k_y=\pi$, the two modes are degenerate at the zone corner.

We can still use the parameters from section \ref{cl}: $t=3.9$, $J=7.25$, $J_4=-0.2J$ and $t'=0$. Then we find the ground state parameter $\sin\theta=-0.34$. The two collective mode dispersions  are plotted in figure \ref{SU2}. In this formalism, the two modes are always degenerate at the zone corner and the experimental results are not well represented.

\section{Inelastic neutron scattering}

We present here the theory for the inelastic neutron scattering from the collective mode of the loop ordered phase. For calculating elastic neutron scattering \cite{he-cmv2}, We approximate the loop current states as 4 spin 1 local moments. Based on the same formalism, we can also compute the inelastic neutron scattering amplitude.

First, we need to express the local moment operator in terms of the generators of Ashkin-Teller model. We can make use of the local moments expression of the 4 loop current states \cite{he-cmv2}. Then the local spin operator at $\vR_1$ can be obtained by $L^i_{\vR_1}=\bra{\psi,\vR_1}L^i\ket{\psi',\vR_1}$, with $\psi,\psi'$ take the 4 possible loop current states.
\begin{widetext}
\be
&&L^x_{\vR_1}=L^x_{\vR_2}=\frac{1}{2}(S^1+T^1),\qquad
L^x_{\vR_3}=L^x_{\vR_4}=-\frac{1}{2}(S^1+T^1),\nonumber\\
&&L^y_{\vR_1}=L^y_{\vR3}=\frac{1}{2}(S^2+T^2),\qquad
L^y_{\vR_2}=L^y_{\vR4}=\frac{1}{2}(S^2-T^2),\nonumber\\
&&L^z_{\vR_1}=-L^z_{\vR3}=\frac{1}{2}(S^3+T^3),\qquad
L^z_{\vR_2}=-L^z_{\vR4}=\frac{1}{2}(S^3-T^3)\nonumber
\ee
Then the inelastic magnetic differential cross section can be expressed in terms of the correlation function of these local moment operators
\be
\frac{d^2\sigma}{d\Omega dE'}&=&\frac{k'}{k}\sum_{\mu\nu}\sum_{a,b}
(\delta_{\mu\nu}-\hat{q}_{\mu}\hat{q}_{\nu})e^{i\vq\cdot(\vR_{a}-\vR_{b})}
\int_{-\infty}^{\infty}\langle L^{\mu}_{-\vq,a}(0)L^{\nu}_{\vq,b}(t)\rangle e^{-i\omega t}dt
\ee
As in the calculation of the collective mode, it is easy to compute the corelation function in terms of Holstein-Primakoff bosons. To the linear order, we find
\be
&&L^x_{\vR_1}=L^x_{\vR_2}=\frac{\cos\theta}{2}(b+\bd+c+\cd),\qquad L^x_{\vR_3}=L^x_{\vR_4}=-\frac{\cos\theta}{2}(b+\bd+c+\cd),\nonumber\\
&&L^y_{\vR_1}=L^y_{\vR3}=\frac{i}{2}(\bd-b+\cd-c),\qquad
L^y_{\vR_2}=L^y_{\vR4}=-\frac{i}{2}(\bd-b+c-\cd),\nonumber\\
&&L^z_{\vR_1}=-L^z_{\vR3}=-\frac{\sin\theta}{2}(b+\bd+c+\cd),\qquad
L^z_{\vR_2}=-L^z_{\vR4}=\frac{\sin\theta}{2}(b+\bd-c-\cd).\nonumber
\ee
Note that only HP boson operator $\bd$ and $\cd$ appear in the above equation and there is no $\dd$. This is consistent with the fact that the local loop current is approximated by spin 1 object. Since $\dd$  flips  $S^z$ by 2, it does not couple to neutron scattering. We can introduce symmetric and anti-symmetric combinations $a_1=(b+c)/\sqrt{2}$ and $a_2=(b-c)/\sqrt{2}$ to simply above equations.  The highest energy mode is a mixture of $\ad_1$ and $\dd$, therefore the highest energy mode should still be observable in principle by neutron scattering with reduced amplitude through exciting $\ad_1$ bosons.

For notational simplicity, we introduce the following quantity.
\be
C^{\mu\nu}=\sum_{a,b}e^{i\vq\cdot(\vR_{a}-\vR_{b})}\int_{-\infty}^{\infty}\langle L^{\mu}_{-\vq,a}(0)L^{\nu}_{\vq,b}(t)\rangle e^{-i\omega t}dt
\ee
Then the cross-section can be written as
\be
\frac{d^2\sigma}{d\Omega dE'}&=&\frac{k'}{k}\Big[(1-\hat{q}_x^2)C^{xx}
+(1-\hat{q}_y^2)C^{yy}+(1-\hat{q}_z^2)C^{zz}\nonumber\\
& &-\hat{q}_x\hat{q}_y(C^{xy}+C^{yx})-\hat{q}_y\hat{q}_z(C^{yz}+C^{zy})
-\hat{q}_x\hat{q}_z(C^{zx}+C^{xz})\Big]
\ee
We find the diagonal elements of $C^{\mu\nu}$ as follows,
\be
&&C^{xx}=2\cos^2\theta(\sin\vq\cdot\vR_1+\sin\vq\cdot\vR_2)^2\nonumber\\
&&\times\int_{-\infty}^{\infty}\Big[\langle a_1(0)a_1(t)\rangle+\langle \ad_1(0)a_1(t)\rangle+\langle a_1(0)\ad_1(t)\rangle
+\langle \ad_1(0)\ad_1(t)\rangle\Big]e^{-i\omega t}dt\nonumber\\
&&C^{yy}=2\cos^2\vq\cdot\vR_1\int_{-\infty}^{\infty}
\Big[\langle a_1(0)a_1(t)\rangle-\langle \ad_1(0)a_1(t)\rangle
-\langle a_1(0)\ad_1(t)\rangle+\langle \ad_1(0)\ad_1(t)\rangle\Big]e^{-i\omega t}dt\nonumber\\
&&+2\cos^2\vq\cdot\vR_2\int_{-\infty}^{\infty}
\Big[\langle a_2(0)a_2(t)\rangle-\langle \ad_2(0)a_2(t)\rangle
-\langle a_2(0)\ad_2(t)\rangle+\langle \ad_2(0)\ad_2(t)\rangle\Big]e^{-i\omega t}dt\nonumber\\
&&C^{zz}=2\sin^2\theta\Bigg(\sin^2\vq\cdot\vR_1\int_{-\infty}^{\infty}
\Big[\langle a_1(0)a_1(t)\rangle+\langle \ad_1(0)a_1(t)\rangle
+\langle a_1(0)\ad_1(t)\rangle+\langle \ad_1(0)\ad_1(t)\rangle\Big]
e^{-i\omega t}dt\nonumber\\
&&+\sin^2\vq\cdot\vR_2\int_{-\infty}^{\infty}
\Big[\langle a_2(0)a_2(t)\rangle+\langle \ad_2(0)a_2(t)\rangle
+\langle a_2(0)\ad_2(t)\rangle+\langle \ad_2(0)\ad_2(t)\rangle\Big]e^{-i\omega t}dt\Bigg)
\ee
Here we have made use of the fact that $\langle a_1a_2\rangle=\langle \ad_1a_2\rangle=\langle a_1\ad_2\rangle=\langle \ad_1\ad_2\rangle=0$.

For the off-diagonal elements, we find that
\be
C^{xy}\propto\int_{-\infty}^{\infty}\Big[\langle a_1(0)a_1(t)\rangle+\langle \ad_1(0)a_1(t)\rangle-\langle a_1(0)\ad_1(t)\rangle
-\langle \ad_1(0)\ad_1(t)\rangle\Big]e^{-i\omega t}dt\nonumber\\
C^{yx}\propto\int_{-\infty}^{\infty}\Big[\langle a_1(0)a_1(t)\rangle-\langle \ad_1(0)a_1(t)\rangle+\langle a_1(0)\ad_1(t)\rangle
-\langle \ad_1(0)\ad_1(t)\rangle\Big]e^{-i\omega t}dt\nonumber
\ee
The cross term of above two equations cancels out. In Eq (\ref{aa}), we show that $\langle a_1a_1\rangle=\langle \ad_1\ad_1\rangle$ and $\langle a_2a_2\rangle=\langle \ad_2\ad_2\rangle$.  Because of this identity, the square terms also cancels out, thus we have $C^{xy}+C^{yx}=0$.

Similarly we find that
\be
C^{zy,yz}&\propto&\Bigg(\int_{-\infty}^{\infty}\Big[\langle a_1(0)a_1(t)\rangle\pm\langle \ad_1(0)a_1(t)\rangle\mp\langle a_1(0)\ad_1(t)\rangle
-\langle \ad_1(0)\ad_1(t)\rangle\Big]e^{-i\omega t}dt\nonumber\\
& &+\int_{-\infty}^{\infty}\Big[\langle a_2(0)a_2(t)\rangle\pm\langle \ad_2(0)a_2(t)\rangle\mp\langle a_2(0)\ad_2(t)\rangle
-\langle \ad_2(0)\ad_2(t)\rangle\Big]e^{-i\omega t}dt\Bigg)\nonumber
\ee
Due to the same reason, we have $C^{zy}+C^{yz}=0$. Then the only non-zero off-diagonal term is
\be
&&C^{xz}=C^{zx}=2\sin\theta\cos\theta(\sin\vq\cdot\vR_1+\sin\vq\cdot\vR_2)
\sin(\vq\cdot\vR_1)\nonumber\\
&&\times\int_{-\infty}^{\infty}\Big[\langle a_1(0)a_1(t)\rangle+\langle \ad_1(0)a_1(t)\rangle+\langle a_1(0)\ad_1(t)\rangle
+\langle \ad_1(0)\ad_1(t)\rangle\Big]e^{-i\omega t}dt
\ee
The HP boson can be expressed in terms of Bogoliubov bosons as follows
\be
&&a_1=u_1\alpha+v_1\alpha^{\dagger}+u_3\gamma+v_3\gamma^{\dagger}\\
&&a_2=u_2\beta+v_2\beta^{\dagger}
\ee
with momentum dependent coefficient $u_i$ and $v_i$ for $i=1,2,3$. Here $\alpha^{\dagger}$, $\beta^{\dagger}$ and $\gamma^{\dagger}$ are the creation operators of the three collective modes. At $T=0$, we find the following correlation functions
\be
&&\int_{-\infty}^{\infty}e^{-i\omega t}dt\langle a_1(0)a_1(t)\rangle
=\int_{-\infty}^{\infty}e^{-i\omega t}dt\langle \ad_1(0)\ad_1(t)\rangle
=u_1v_1\delta(\omega-\omega_1(\vq))
+u_3v_3\delta(\omega-\omega_3(\vq))\nonumber\\
&&\int_{-\infty}^{\infty}e^{-i\omega t}dt
\langle a_1(0)\ad_1(t)\rangle=u_1^2\delta(\omega-\omega_1(\vq))
+u_3^2\delta(\omega-\omega_3(\vq))\nonumber\\
&&\int_{-\infty}^{\infty}e^{-i\omega t}dt
\langle \ad_1(0)a_1(t)\rangle=v_1^2\delta(\omega-\omega_1(\vq))
+v_3^2\delta(\omega-\omega_3(\vq))\nonumber\\
&&\int_{-\infty}^{\infty}e^{-i\omega t}dt
\langle a_2(0)a_2(t)\rangle=\int_{-\infty}^{\infty}e^{-i\omega t}dt\langle \ad_2(0)\ad_2(t)\rangle=u_2v_2\delta(\omega-\omega_2(\vq))\nonumber\\
&&\int_{-\infty}^{\infty}e^{-i\omega t}dt
\langle a_2(0)\ad_2(t)\rangle=u_2^2\delta(\omega-\omega_2(\vq))\nonumber\\
&&\int_{-\infty}^{\infty}e^{-i\omega t}dt
\langle \ad_2(0)a_2(t)\rangle=v_2^2\delta(\omega-\omega_2(\vq))
\label{aa}
\ee
Since we ignored the damping effect, there are only delta function like peaks. Note all 3 collective mode dispersions appear in the above correlation functions. Collecting all results, we find
\be
\frac{d^2\sigma}{d\Omega dE'}&=&\frac{k'}{k}\Bigg(\Big[4(1-\hat{q}_x^2)F_{xx}
+(1-\hat{q}_z^2)F_{zz1}-\hat{q}_x\hat{q}_zF_{xz}\Big]\nonumber\\
& &\times\Big[(u_1+v_1)^2\delta(\omega-\omega_1(\vq))
+(u_3+v_3)^2\delta(\omega-\omega_3(\vq))\Big]\nonumber\\
& &+(1-\hat{q}_y^2)F_{yy1}\Big[(u_1-v_1)^2\delta(\omega-\omega_1(\vq))
+(u_3-v_3)^2\delta(\omega-\omega_3(\vq))\Big]\nonumber\\
& &+[(1-\hat{q}_y^2)F_{yy2}(u_2-v_2)^2
+(1-\hat{q}_z^2)F_{zz2}(u_2-v_2)^2]\delta(\omega-\omega_2(\vq))\Bigg)
\ee
with $F_{xx}=2\cos^2\theta(\sin\vq\cdot\vR_1+\sin\vq\cdot\vR_2)^2$, $F_{yy1}=2\cos^2(\vq\cdot\vR_1)$, $F_{yy2}=2\cos^2(\vq\cdot\vR_2)$, $F_{zz1}=2\sin^2\theta\sin^2(\vq\cdot\vR_1)$, $F_{zz2}=2\sin^2\theta\sin^2(\vq\cdot\vR_2)$ and $F_{xz}=2\sin\theta\cos\theta(\sin\vq\cdot\vR_1+\sin\vq\cdot\vR_2)
\sin(\vq\cdot\vR_1)$.\\
\end{widetext}


\begin{thebibliography}{99}
\bibitem{note}
Actually, since the interactions between ${\bf L}$ at different sites are due to currents and not exchange, the model should have two parameters  $J_{21}$ and $J_{22}$ with the following restriction: The symmetry of the problem  requires that the interactions among the $\sigma$ operators for two sites in the $(x,y)$-direction with respect to each other equals the interaction between the $\tau$ operators in the $(y,x)$ directions \cite{sudbo-2}. We have ignored this subtlety below by taking $J_1=J_2 = J_{21} = J_{22}$ for most of the calculations in the paper.
\end{thebibliography}
\end{document}